# Rigorous Safety Analysis and Design of ADAS and ADS – Implications on Tools

Juan Pimentel

Principal Consultant

Omnex

**Abstract**

Currently, a major concern is the insufficient level of safety offered by commercial automated vehicles and/or services such self-driving vehicles, self-driving trucks, and robotaxis. Unfortunately, stakeholders do not agree on definitions and characterizations of what is meant by safety of automated vehicles including how to measure it and how to design for it. This paper sheds some light into the answers to important questions about the safety of automated vehicles. In addition, we identify rigor as a significant missing requirement in the current literature, we also provide a discussion of rigor in the design, development, and commercialization of automated vehicles. Furthermore, we discuss software tool requirements at the organizational level to support a rigorous approach for the analysis, design, and commercialization of automated vehicles. An ALM tool, EwQIMS, is introduced emphasizing its rigorous features of its functional safety module that implements much of the ISO 26262 standard.

## 1. Introduction

Although with some setbacks, the automotive industry continues to develop automated vehicles at various automation levels and these include as self-driving vehicles, robotaxis, ADAS, ADS and the like. Recently, there has been new developments as well as drawbacks in the industry most notable one by Cruise and another by Tesla. Cruise lost its license to operate in SF due to an accident where a driverless Cruise car hit a pedestrian, stopped, and then tried to pull over, dragging the pedestrian approximately 20 feet[1] Tesla on the other hand seems to have difficulties in its realization of self-driving vehicles and its FSD system seems to be stuck on beta, at a level 2 of the SAE automation system[2]. The entire automotive industry was overly optimistic just a few years ago, with expectations that self-driving vehicles was to be widely available as early as 2018[3]. The industry spent billions of dollars, some companies ceased to exist, most notably Argo.ai, which was backed by Ford.

Currently, the state of the industry of automated vehicles is uncertain. Tesla just announced that on August 8, 2024 it will unveil its robotaxi and there is a great deal of discussion as to what exactly will be unveiled on such a date[4]. Why all of this uncertainty? Surely, one of the main reasons is that designing, implementing, and commercializing automated vehicles turned out to be much more difficult than anticipated. The automated vehicle industry has retreated now, by adopting a more cautious approach and beginning to offer ADAS systems operating at levels 1 and 2 at the beginning and gradually offer more automated systems, i.e., ADS, which would include levels 3 and 4 at a later time. The so-called, level 5 has an indefinite timetable, and some analysts predict that it will never be accomplished[5].

---

[1] https://www.wired.com/story/cruise-robotaxi-self-driving-permit-revoked-california/
[2] https://www.thestreet.com/technology/whats-stopping-tesla-from-achieving-level-3-self-driving
[3] https://jalopnik.com/2018-was-a-hard-reality-check-for-autonomous-cars-1831182272
[4] https://www.theverge.com/2024/4/5/24122384/tesla-robotaxi-reveal-date-elon-musk-august-8

Chief among the concerns voiced by shareholders is the insufficient level of safety offered by commercial automated vehicles and/or services such self-driving vehicles, self-driving trucks, and robotaxis[6]. An additional concern is the low acceptance of automated vehicles from the general public and also the prevalence of social media that emphasizes anything even remotely negative for automated driving, failing to mention how 90+% of all accidents are caused by human error. Thus, an important contemporary goal for vehicle manufacturers is to design, implement, test, and commercialize automated vehicles with a sufficient level of safety that can be accepted by all stakeholders, particularly the general public and governments at all levels.

Many OEMs sell vehicles today offering functionality that belongs to ADAS, at levels 1 and 2, some even classify some available commercial vehicles as Level 2+ and level 2++ meaning that the functionality offered is beyond level 2 but not sufficient enough to make it to level 3. Although similar in functionality, the various ADAS offered by OEMS are not exactly alike in the way they are conceived, designed, implemented, tested, and commercialized. On the one hand we have Tesla and few others which is following a unique methodology using vision only and an end-to-end approach to perception, planning, and control. On the other hand, we have the rest of the industry which have opted for a more traditional approach to ADS, that covers levels 3, 4, and 5, rooted in product lifecycle, risk based, and systems engineering. So far, neither approach have proven to be more effective than the other in offering solutions to the difficult problem of developing and commercializing automated vehicles with a sufficient level of safety.

## 2. Current Approaches to ADS/ADS Design

There are basically two main approaches that are being currently used in the design of ADAS/ADS: a traditional one which is based on risk assessment, system engineering, and procedural programming and a second approach which is based on vision only, neural networks based, and a closed loop, end-to-end integration of perception, planning, and control.

**A) Approach based on safety risk, systems engineering, and procedural programming.**

This the traditional approach based on decades of experience in designing safety critical systems in other industry segments such as nuclear, avionics, aerospace, continuous processes, manufacturing and others. The application of this approach to the design of ADAS/ADS is based on state of the art of the underlying technologies and is exemplified in recent standards such as ISO 26262, ISO 21448, ISO 34502, ISO 34503, ISO 5083, and others.

The traditional approach is based on a framework that includes the following sub-systems: perception, planning, control, and actuation. The perception sub-system includes an array of perception sensors that include ultrasonics, cameras, radars, LIDARs, GPS, GNSS, and others. Typically, a sensor fusion module integrates the sensor outputs for some of these sensors to produce value added information. The perception system also includes models of the driving environment. The planning sub-system generates a path or trajectory for the vehicle to follow based on an environmental model of the scenes and situations. The control sub-system generates the longitudinal and lateral controls based on the commanded trajectories obeying driving laws and traffic signals while avoiding collisions with other vehicles, pedestrians, and other objects. The actuation sub-system converts the control signals into actual physical lateral and longitudinal movement of the vehicle.

---

[5] https://www.carriermanagement.com/features/2023/09/22/253499.htm
[6] https://www.forbes.com/advisor/legal/auto-accident/perception-of-self-driving-cars/

To reduce complexity, the traditional approach uses HD maps and geofencing by restricting the driving environment to specific areas called the operational design domains (ODD) [14]. Neural networks are typically used in the various sub-systems but not as an end-to-end fashion as the visual only approach to be described in section 1.B below. An example of a system using this approach is *Waymo Driver*, an ADS level 4 technology developed by Waymo [5].

**Waymo Driver**

To learn many tasks associated with the driving functions, *Waymo Driver* uses machine learning (ML) techniques based on deep neural networks. In particular, ML enables Waymo to detect and recognize all kinds of objects with high accuracy in all conditions, understand and interpret complex object and scene semantics, and make vehicle driving smooth and human-like[7]. One advantage of the ML methodology is that it adapts to new challenges and learns from experience fast. That, combined with robust training and evaluation infrastructure, allows Waymo to rapidly and regularly enhance the *Waymo Driver*'s capabilities. Vehicle software updates are prepared and delivered quite periodically. Each update undergoes rigorous testing and often unlocks new features and performance improvements for the Driver across an entire fleet of vehicles — from parking and multi-point turn maneuvers to handling heavier levels of fog and rain — as if a driving lesson for one teaches the whole fleet. This allows us to safely deploy not just individual cars but our entire fleet in increasingly more complex driving environments and expand to new cities faster. Waymo Driver's automated driving features have even been developed around the use of "operational design domain" ("ODD") — e.g., geo-fences, reflecting the industry's view that imposing limits on the ODD will be an important step in developing autonomous systems that can operate above Level 2.

**B) Vision only, Neural network based, End to End Approach**

This approach, primarily being championed by Tesla, is relatively new which was made possible by recent advances in computer vision, deep neural networks, and AI techniques, all working in a closed loop fashion[8]. This vision only approach relies only on cameras and do not use other typical perception sensors such as ultrasonic, radar, LIDAR, and GPS. For example, in 2021, Tesla announced that it would also be discontinuing its use of radar sensors, a device that provides autonomous vehicle systems with information about the distance and relative speed of objects around a vehicle[9]. Radar sensors are used by all other major developers working on high levels of vehicle autonomy.

---

[7] https://waymo.com/blog/2023/08/the-waymo-drivers-rapid-learning-curve/

[8] https://www.engineering.com/story/now-revealed-why-teslas-have-only-camera-based-vision
[9] https://www.tesla.com/support/transitioning-tesla-vision

At the Tesla's 2019 Autonomy Investor Day, the Company has publicly stated that it will develop its autonomous driving system without the use of LIDAR[10]. LIDAR is a technology, used by other companies attempting to develop high levels of vehicle autonomy, that can detect distance by measuring the time it takes beams of light to be transmitted and then returned to a sensor. While Tesla contends that it can determine distance entirely through ordinary passive optical data, and while this may be technically possible, the decision to reject this technology increases the challenge of developing a safe and reliable system. Importantly, it reduces the availability of multiple and independent sources of data that can be used to ensure redundancy and reliability.

In addition to rejecting these specific sensors, the vision only approach also foregoes other technologies and methodologies commonly used in the industry, such as HD maps and geofencing based on ODDs. The logic of the end-to-end approach is to collect data from a fleet of vehicles driven by its owners, feed that data to an off-line system to train its neural network, then deploy the trained neural network on the vehicle, and then iteratively improve the system based on the feedback loop between the collected data and vehicle performance [1]. This approach yields results but is slow moving because improvements rely on the process of slowly collecting data about perceived issues and then iteratively making changes to improve performance. Nevertheless, it is not guaranteed that this approach will address all of the detected safety issues.

The iterative machine learning based approach makes it exceptionally difficult to reliably predict progress. A system may make improvements for a time, before reaching a point where progress stalls. This phenomenon is well-illustrated by Tesla's rate of progress in developing improvements in its technology. For example, in Tesla's September 22, 2020 annual meeting[11], it was disclosed that Tesla was substantially changing its approach to providing data to its neural network system – moving from labeling individual images to video segments – because "a couple years ago" the company had become "kind of stuck in a local maximum," meaning that the system was "improving, but like the improvements kind of started tailing off and just not getting where they needed to be . . . so the company had to do a fundamental rewrite of the entire Autopilot software stack and all of the labeling software as well." The prospect of reaching "local maximum(s)" is heightened by an approach that relies heavily on feeding data to machine learning algorithms and waiting for improvements. To date, it has not been proven in the research and development community that an automated driving system based entirely on neural networks has the sufficient level of functional safety to be deployed in actual systems.

Initially, Tesla used neural networks primarily for the perception system involving the detection of vehicles, traffic signs, pedestrians, lane lines, and other objects. For planning and control, Tesla used traditional procedural programming that resulted in planning and control software having around 300,000 lines of code. Around January 20, 2023, Tesla moved to a mostly neural network approach to use neural networks for planning and control. On August 1, 2023, Tesla announced that this decision resulted in a reduction of procedural code from 300,000 lines of code to about 3,000 lines of code[12], a reduction of 100 times smaller. In other words, its current system is essentially entirely dependent on the neural network trained system to produce accurate results in terms of perception, planning, and actual driving decision making, and not on "hard coded" procedural rules (e.g., when red light detected, slow to a stop, etc.)

---

[10] https://www.washingtonpost.com/technology/2019/07/17/tesla-floats-fully-self-driving-cars-soon-this-year-many-are-worried-about-what-that-will-unleash/
[11] https://www.rev.com/blog/transcripts/tesla-2020-battery-day-transcript-september-22
[12] https://www.nextbigfuture.com/2023/08/tesla-fsd-is-all-ai-without-hardcoding.html

In summary, the end-to-end, vision only approach uses a pre-trained deep neural network to generate vehicle acceleration, braking and steering controls directly from images and videos, i.e., in an end-to-end fashion, collected by a fleet of vehicles. This end-to-end, vision only approach foregoes other sensors, technologies, and methodologies such as ultrasonics, radar, LIDAR, HD maps, and geofencing based on ODDs.

## 3. Main Issue

One significant issue with the current developments and commercialization of automated vehicles by developers of automated vehicles is summarized in the following question: *In spite of over 10 years of intensive investment, research, development, and commercialization by many stakeholders, why don't we have commercially available automated vehicles with a sufficiently accepted level of safety?*

Although we do not claim to have a definitive answer to the above question, in this paper we formulate additional related questions while discussing elements of some answers to these questions and in the process, we identify a significant requirement that will enable the achievement of automated vehicles with a sufficient level of safety.

The additional questions that will help addressing the main issue described above are:

1. What is safety and more specifically what is automated vehicle safety?
2. How do you know if an ADS/ADS is safe? How do you measure automated vehicle safety? How safe is safe enough?
3. How do you design and deploy safe automated vehicles? What are the main methodologies of? What else is needed?
4. What is rigor in the development, implementation, and testing of ADAS/ADS systems?
5. What is the role of a rigorous safety analysis and design approach? Where is rigor needed?
6. What kind of tools support rigorous safety analysis and design for automated vehicle development? What kind of tool categories are out there?

Currently, the automated vehicle industry, including all of its stakeholders are addressing some of these questions. However, there are no clear answers nor there is a consensus on some of the answers being proposed or discussed by a multitude of stakeholders. The contributions of this paper are: a) to shed additional light into the answers to these questions, b) the identification of rigor as a significant requirement to help answer question 3, c) a discussion of rigor in the design, development, and commercialization of automated vehicles, and d) a discussion of software tool requirements at the organizational level to support a rigorous approach for the analysis, design, and commercialization of automated vehicles.

In the following we address the above questions.

## 4. Question 1 – What is safety and more specifically what is automated vehicle safety?

Some endeavors are not safety-critical, meaning that any component failure of the systems or any unintended behavior do not lead to any hazards resulting in any sort of harm or accident affecting human beings, and even if they do, their impact is minimal so that they can be tolerated. For example, software systems dealing with convenience, user preferences, human-machine interface (HMI), enabling/disabling of convenience features, and the like are primarily non-safety critical. Many aspects of automated vehicles are safety critical [2], involving primarily three types of hazards and associated safety risks: architectural, behavioral, and in-service operational hazards [5]. Thus, there are several aspects or types of automated vehicle safety. One important aspect is that of "functional safety" related to architectural aspects as addressed by the standard ISO 26262 [15]. Another aspect is that of "the safety of the intended function" (SOTIF) as addressed by the standard ISO 21448 [16] that includes aspects related to driving rules or driving policy that constitutes the "behavioral" aspect of safety. Still another aspect is related to the safety of deployed automated vehicles when they "operate", interact, and sometimes interfere with other services in a city environment, e.g., firefighters, ambulances, or police. Waymo calls the latter two safety aspects a *Behavioral* and *In-service Operational* [5]

There is no doubt that safety is paramount for the development and deployment of automated vehicles [1]. Unfortunately, there is no agreement from the multiple stakeholders as to the definition and the multiple elements of what constitute safety in the context of automated vehicles [1, 11, 12 18]. The various stakeholders such as the general public, governments at the country, state, province or city level, the law, public service organizations, standard bodies, insurance companies, OEMs, vehicle suppliers, research organizations, and academia each have their own view of what constitute safety and they do not necessarily agree. In some communities, this question is known as how safe is safe enough [21].

Safety, and in particular automated driving safety have been defined in many different ways. For the purpose of this paper, we adopt the following definition of automated driving safety as the *absence of unreasonable safety risk*, which is in accordance with industry standards [5].

## 5. Question 2 - How do you know if an ADS/ADS is safe? How do you measure automated vehicle safety? How safe is safe enough?

An ADS/ADS is deemed safe if an overall positive level of safety or safety score can be determined based on a number of well accepted safety metrics. In terms of measuring the level of safety, there are many metrics proposed or suggested by various stakeholders ranging from researchers, vehicle manufacturers, the general public, to research organizations, to insurance organizations and others. An early research paper by Kalra and Paddock [3] showed that fully autonomous vehicles would have to be driven hundreds of millions of miles and sometimes hundreds of billions of miles to demonstrate their reliability in terms of fatalities and injuries. One reason for the estimate given by the authors is that current traffic fatalities and injuries are probabilistic rare events compared to vehicle miles traveled.

The RAND corporation has developed a set of metrics [4], called measures, for measuring automated vehicle safety in three categories: a) Standards, processes, procedures, and design, b) Leading measures: Infractions, roadmanship, and disengagements, and c) Lagging measure involving actual harm, crashes, accidents and their outcomes. Table 1 provides a format for all the possible outputs of the RAND safety metrics in the context of product lifecycle state and test setting.

Table 1. RAND safety metrics for measuring automated vehicle safety.

| Stage | Setting | Leading Measures | | | Lagging Measures |
|---|---|---|---|---|---|
| Development Demonstration Deployment | Simulation Closed course Public roads | Infractions | Roadmanship | Disengagement | Outcome |

As another example, the Insurance Institute for Highway Safety (IIHS)[13] considers the following safety measures for assessing "partial automation safeguards" pertaining to an SAE level 2: a) monitors both the driver's gaze and hand position b) Uses multiple types of rapidly escalating alerts to get driver's attention, c) Fail-safe procedure slows vehicle, notifies manufacturer and keeps automation off limits for remainder of drive, d) Automated lane changes must be initiated or confirmed by the driver, e) Adaptive cruise control does not automatically resume after a lengthy stop or if the driver is not looking at the road, f) Lane centering does not discourage steering by driver, g) Automation features cannot be used with seat belt unfastened, and h) Automation features cannot be used with automatic emergency braking or lane departure prevention/warning disabled. In early March, 2024 the IIHS tested 14 systems according to the above measures and only one earned an acceptable rating; two were rated marginal, and 11 were rated poor[14].

Koopman and Widen [10] argue that acceptably safe AVs must satisfy criteria along multiple different dimensions simultaneously. These criteria include a) Achieving a Positive Risk Balance (PRB) for comparable conditions, b) Mitigating risk transfer onto vulnerable populations, c) Avoiding negligent computer driver behavior, d) Conforming to industry consensus safety standards, e) Meeting regulatory requirements for risk on a fine-grain basis, and f) Addressing ethical and equity concerns. One can think of safety metrics designed aligned each of the above criteria. The notion of Positive Risk Balance (PRB) was stated in a report by the BMVI Ethics Commission (BMVI 2017) as follows, "licensing of automated systems is not justifiable unless it promises to produce at least a diminution in harm compared with human driving, in other words a positive balance of risks." Regarding risk transfer onto vulnerable populations, including risk inequalities, the EC Ethics Report (2020) recommends that no category of road user should be at risk of increased harm, even if net harm to all road users has been reduced. Maintaining a great driving record in complex driving environment is difficult for human drivers and much more so for computer drivers. Avoiding negligent computer driver behavior entails holding computer drivers to the same standards of negligent and reckless driving behavior that already apply to human drivers. Most stakeholders agree and place emphasis on the importance of conforming to industry consensus safety standards and government regulations in the design and development stages to achieve safe AVs. Such safety standards include ISO 26262, ISO 21448, UL 4600 [6,15,16], among many others. Among regulations we have some from the NHTSA and some from the UNECE such as regulations 157, 153, and 152. Finally, there are ethical and equity concerns that must be addressed because they can directly affect and constrain design choices, activities, and deployment decisions that are relevant to the safety of AVs. Examples include allowing developmental testing of potentially defective prototype vehicles in areas where the public did not consent or in historically disadvantaged areas.

---

[13] https://www.iihs.org/ratings/partial-automation-safeguards
[14] https://www.iihs.org/news/detail/first-partial-driving-automation-safeguard-ratings-show-industry-has-work-to-do

Tesla has introduced *Safety Score* [Beta] to assess driving behavior of Tesla vehicles with Autopilot software enabled based on several metrics called Safety Factors[15]. These are combined to estimate the likelihood that driving behavior could result in a future collision. Safety Score Beta is intended to provide drivers transparency and feedback of their driving behaviors to encourage safer driving and potentially pay less for their insurance. The Safety Factors include a) Forward Collision Warnings per 1,000 Non-Autopilot Miles, b) Hard Braking, c) Aggressive Turning, d) Unsafe Following, e) Excessive Speeding, f) Late-Night Driving (11 PM - 4 AM), g) Forced Autopilot Disengagement, and h) Unbuckled Driving.

Finally, some purely empirical lagging metrics, such as *number of miles per disengagement*[16] and number *of accidents per miles driven* are being used by some of the public.

In summary, it is not straightforward to characterize the safety of automated vehicles as there are many stakeholders each viewing safety from a different perspective and the same thing can be said about measuring the level of safety of deployed automated vehicles. Accordingly, there are many methodologies proposed and being used for the design and test of automated vehicles that cater to the multiple stakeholders. Unfortunately, there is no universal agreement among the various stakeholders as to the set of safety metrics and how to determine if an automated vehicle is sufficiently safe or not.

## 6. Question 3 - How do you design and deploy safe automated vehicles? What are the main methodologies? What else is needed?

Much effort has been placed by many stakeholders on developing approaches and methodologies that would enable automated vehicles with a sufficient level of safety, thus assuring the safety of automated vehicles. Unfortunately, such efforts have not yielded good results, otherwise safe automated vehicles would be a reality today. One reason for the lack of real and significant breakthroughs in this area is the fact that achieving the necessary functionality together with the required dependability for automation levels 3, 4, and 5 is particularly challenging [11]. Stakeholders, particularly OEMs are not in total agreement as to the way to design and deploy safe automated vehicles. Thus, there is not a lack of approaches and methodologies to consider. Clearly, there must be holes in the proposed approaches or missing components that must be identified. It is beyond the scope of this paper to review and analyze the available approaches and methodologies that have been proposed or suggested by various stakeholders. Nevertheless, we will review only a few in the following.

Burton et al. [7] proposed an engineering, ethical, and legal perspective for assuring the safety of automated vehicles in terms of semantic gaps, responsibility gaps, and liability gaps respectively. The semantic gap arises because the normal conditions are not met for manufacturers, particularly designers, to provide a complete specification of the system. The semantic gap represents the difference between the implicit intentions on the system's functionality and the explicit, concrete specification that is used to build the system. It is a risk in the design phase, but it is also an ongoing problem, with amendments to specification being required after implementation and deployment [7]. The responsibility and liability gaps extend this notion to deserve moral blame for the automated vehicle's decisions and to be liable to pay compensation to those injured by an automated vehicle respectively. The authors identify three root causes for the gaps namely, a) the complexity and unpredictability of the operational domain. Automated vehicles typically operate within an environment that cannot be fully specified at design time, b) the inherent complexity and unpredictability of the system itself due to computational techniques used and continual change through interactions within the domain, and c) the increasing transfer of decision function to the system, whereby the human actor is either replaced completely or relieved of a substantial cognitive load.

---
[15] https://www.tesla.com/support/insurance/safety-score#version-2.1 Unsafe Following
[16] https://teslamotorsclub.com/tmc/threads/personal-fsd-beta-disengagement-rate.278493/

The authors provide the following suggestions for reducing the semantic, responsibility, and liability gaps [7]:

1. Limiting functionality to well-defined scenarios [13] or ODDs for which a clear understanding of the safety risks and system capabilities already exist or can be developed.
2. Limiting the deployment of ML algorithms to well-defined and constrained functionality with restricted safety impact
3. Reducing the delegation of decision function to the system. This can take a number of forms, including the requirement for driver supervision at Level 3 of highway driving automation.
4. Detailed system design and evaluation where safety-critical scenarios are identified during the system design phase.
5. Post-hoc analysis and allocation of blame and liability
6. Development of a safety case that will stand up legally upon deposition in a court of law.
7. Monitoring and dynamic assurance - due to the complexity and uncertainty of the system and its operational environment, e.g., using ML techniques [9].
8. Soft law, e.g., an industry-wide code of practice, which companies adopt.
9. Regulation and governance.

**A. Dependability Considerations**

When analyzing ADAS/ADS technologies, it is well accepted in scientific and engineering communities that such technologies can be productively analyzed by considering their Functionality, Operability, Dependability, and Performance, which terms have well understood meanings within various engineering communities. Within the topic of "dependability," there are subcategories of safety, security, reliability, and availability. Functionality is about what functions the vehicle performs at a particular level of driving automation, e.g., highway intelligent cruise control at automation level 2, or complete independent operation on all roadways at automation level 5. Additional examples of functionality include lane keeping, lane change, intersection driving, unprotected left turns, parking, etc. Operability is about the functioning of the vehicle as a component of a complex system made up of other vehicles, pedestrians, bicyclists and belonging to a wide range of stakeholders, e.g., the vehicle must be able to safely operate in construction zones, the vehicle must co-exist with first responders such as fire trucks, emergency vehicles, police, etc. Waymo refers to this as in-service operational risk [5] Dependability is the overall reliance that is placed on the vehicle to provide such functionality and operability. There are four attributes of dependability: safety, security, reliability, and availability. Performance involves the manner, quality, and efficiency of functional and operational aspects of driving.

An IFIP (International Federation for Information Processing) working group performed a multi-year project on highly dependable and secure operation of intelligent vehicles, verified and validated with respect to strict dependability (particularly safety) and security requirements by rigorous state-of-the-art methods. Principal findings of the project point to significant shortfalls in technologies, cost, governance, and societal aspects in achieving the end goal of safe and secure self-driving intelligent vehicles, generally referred to as SAE Level 4 or 5 [11]. Self-driving at L4 is subject to limited conditions specified by an Operational Design Domain (ODD). With sufficiently restrictive ODDs, good progress is being made toward achieving this goal. On the other hand, at L5 an intelligent vehicle can drive anywhere under any conditions. Widespread deployment of such vehicles on public roads, much hyped at the outset of the project, appears to have receded into the more distant future [11].

One can think of safety methodologies aligned with each of the safety criteria developed by Koopman and Widen [10] namely: a) Achieving a Positive Risk Balance (PRB) for comparable conditions, b) Mitigating risk transfer onto vulnerable populations, c) Avoiding negligent computer driver behavior, d) Conforming to industry consensus safety standards, e) Meeting regulatory requirements for risk on a fine-grain basis, and f) Addressing ethical and equity concerns.

**B. Development of a Safey case**

As suggested by Burton et al. [7], developing a safety case is important for reducing the gaps to achieve an acceptable level of safety for automated vehicles. Waymo also places a great deal of importance to the development of a safety case [5]. Waymo uses the definition of a safety case from the UL 4600 standard stating: *"A structured argument, supported by a body of evidence that provides a compelling, comprehensible, and valid case that a system is safe for a given application in a given environment"* [5, 6]. However, unlike traditional emphasis placed on the formatting aspects of a safety case[17], Waymo emphasizes building a credible safety case based on strong arguments rather than just documenting such arguments and corresponding evidence. In this context, Waymo introduces the concept of *(Safety) Case Credibility Assessment* (CCA) consisting of two ingredients, a) credibility of the argument, and b) credibility of the evidence. In turn, credibility of the argument includes the assessment of acceptance criteria reasonableness and the assessment of argument suitability. Credibility of the evidence includes a confidence assessment and a coverage assessment [5].

**C. Other Methodologies Based on Industry Consensus**

So far, while addressing question 3 above, we discussed ADAS/ADS design methodologies suggested by specific authors or projects. However, as also noted, there is a wealth of methodologies that are being used by industry consensus that includes for example those incorporated in standards and regulations. In the following, in no particular order, we briefly summarize complementary design methodologies to achieve safe automated vehicles. Nevertheless, the addressing of methodologies is not exhaustive as it is beyond the scope of this paper.

**a) Systems Engineering and the V Model**

Systems Engineering is a methodology widely adopted in automotive design and other endeavors. Basically, it consists of viewing a system or endeavor as an interconnected set of components making up the whole which are further decomposed into sub systems at lower levels of abstraction until reaching a level where the system constituent components can be realized or implemented into hardware, software, or a combination thereof. Systems engineering work begins at the top and most abstract level and proceeds iteratively by decomposing functionality and design requirements into more detailed lower levels. To cater for product development, systems engineering uses a V model where activities such as the development of functions, requirements, and specifications are on the left side of V and activities such as implementation, integration, testing, verification and validation are on the right side of the V. The main tenets of system engineering include functions, requirements, architecture, requirements allocation, design specifications, product lifecycle, V model, testing, verification, and validation.

---

[17] https://misra.org.uk/product/misra-gasa/

**b) More Formal, Quantitative, and Stringent Methodologies**

References [5, 11] among others suggest more formal, quantitative, and stringent analysis and design methodologies and much of these have been incorporated in standards, regulations, and technical reports. More specifically, [11] recommends that quantitative standards of safety must also be developed, to complement the qualitative ISO 21448: Safety of The Intended Functionality (SOTIF) and other standards. These would provide the community a common measurement yardstick: a specific design criterion for the industry, an objective standard to certify vehicles for the regulators, and a specific goal for technology developers and something to aim for by the research community.

**c) New Product Design and Product Lifecycle**

By its very nature, automated vehicles call for new product designs which are typically based on a product lifecycle. Product lifecycle consists of viewing the entire product development as being composed of various phases beginning with planning, concept design, and design development including integration, testing, verification, and validation. A final phase called post-development phase completes the product lifecycle. The V model plays a crucial role at the concept design and the design development phases. Well known risk-based methodologies for functional safety have been integrated into product lifecycle phases; such is the case with the standards such as ISO 26262 and ISO 21448. The very nature of an automotive product lends itself to cater to an iterative development lifecycle that includes a concept phase, product development phase, and a post development phase all performed in an iterative fashion.

**d) international standards**

The development and availability of standards, such as ISO 26262, and ISO 21448, have proven crucial for efficient designs and developments of new automotive products involving safety critical sub-systems [8]. These standards have adopted premises from systems engineering in an effort to deal with complexity. These premises are embedded in various aspects of the standards that include design, development, processes, and management. The above standards are risk based, follow system engineering principles, and offer a product lifecycle view. For example, ISO 26262 begins a design by developing an Item Definition in the concept phase which is analyzed for safety hazards. A HARA is performed next to select significant safety hazards which are then turned into safety goals in order to select appropriate risk treatment options. The process then continues with deriving design requirements that would realize (i.e., implement) the safety goals. Such design requirements are then implemented in appropriate hardware and/or software components of the electrical/electronic system architecture. The design and implementation activities are performed at a system level first and then at the hardware and software levels. The above process is performed iteratively by refining the architecture, requirements, and design specifications in each iterative cycle.

**e) Well accepted testing, verification, and validation techniques**

Once components and sub-systems are being implemented, integrated, and assembled into the target product a great deal of testing, verification, and validation activities are required. Compliance standards already require these and a cadre of commercial tools are available to support these activities[18]. Testing, verification, and validation techniques are specified based on appropriate acceptance criteria and validation targets defined at the analysis phase [16].

Although there is a wealth of methodologies to design and deploy automated vehicles, the degree of developments is by no means complete; there is a significant missing component. In this paper we argue that the important missing component is rigor in all aspects of development and deployment, particularly at the analysis and design phases. While addressing question 4 below, we clarify what we mean by rigor.

## 7. Question 4 - What is rigor in the analysis, development, implementation, and testing of ADAS/ADS systems? Why do we need rigor in a safety analysis and design approach?

Regardless of how automated vehicle safety is defined, how it is measured, and how it is designed into automated vehicles, one thing is clear: a rigorous approach must be used in all aspects of automated vehicle safety analysis, design, and development. This follows from the wealth of research, developments, and experience from similar safety-critical systems in other industries such as nuclear, avionics, aerospace, process control, etc. [11] Indeed, a rigorous safety analysis and design approach in avionics for example has enabled the industry to design, manufacture, and commercialize aircraft with extremely low accident rates and well accepted by the general public [12]. Thus, it is of the outmost importance to follow a rigorous safety analysis and design approach in the design of automated vehicles with a high level of safety. But what constitutes a rigorous safety analysis and design approach? We address this question below.

Automated vehicles are extremely complex, and this complexity extends to its safety analysis and design [7]. There is an established body of knowledge about rigorous safety analysis and design in various industry domain such as nuclear, avionics, process control, and manufacturing [11]. An equivalent body of knowledge for automated vehicles is just emerging but it is not mature enough when compared to the aforementioned industries; we argue that the missing component is a rigorous approach that includes quantitative acceptance criteria and validation targets. We contend that a rigorous safety analysis and design methodology is paramount for ensuring the safety of automated vehicles.

One can use a wealth of safety approaches and methodologies and still deliver systems with an insufficient level of safety. In this context, what we mean by rigor *is to approach and develop systems in an extremely detailed fashion in a manner that is precise, exact, accurate, clear, compelling, comprehensive, valid, exhaustive, stringent, complete, holistic, sufficiently covered, and robust*. To exhibit the above characteristics, rigor can benefit from both, qualitative and quantitative approaches. Rigor in analysis and development is a necessary ingredient to achieve a sufficient level of safety of in the design of automated vehicles. While it is difficult to provide a clear and precise definition of rigor, the following examples illustrates some of its attributes.

---

[18] https://en.wikipedia.org/wiki/List_of_requirements_engineering_tools

***A. Example 1: Rigor in the development of an Item Definition***. The functional safety standard ISO 26262 begins its lifecycle development with an Item Definition, in the Concept Phase (Part 3 of the standard). The Item Definition is crucial to the remaining tasks not only in the Concept Phase but also in the Product Design and Development Phase (Parts 4, 5, and 6) and Post development Phases (Part 7). ISO 26262 identifies a few essential pieces of information that "should be included" in the Item definition and they are: functionality, interfaces, environmental conditions, legal requirements and hazards. In addition, the standard requires the following requirements of the item to be made available [15]:

> a) legal requirements, national and international standards;
>
> b) the functional behaviour at the vehicle level, including the operating modes or states;
>
> c) the required quality, performance and availability of the functionality, if applicable;
>
> d) constraints regarding the item such as functional dependencies, dependencies on other items, and the operating environment;
>
> e) potential consequences of behavioural shortfalls including known failure modes and hazards, if any; and
>
> f) the capabilities of the actuators, or their assumed capabilities.

The more items in the above list are included in the Item Definition, the more rigor in its characterization. As an example of an added rigor, Fig. 1 depicts the state transition diagram of the highway assist (HWA) feature of an ADAS system performing at level 2 of the SAE automation hierarchy. Other ADAS features depicted in Fig. 1 are Intelligent cruise control (ICC) Automated Lane change (ALC), Lake keeping assist (LKA), and Autonomous emergency braking (AEB).

In terms of the boundary of the item, ISO 26262 requires that its interfaces, and the assumptions concerning its interaction with other items and elements, be defined considering:

> a) the elements of the item.
>
> b) the assumptions concerning the effects of the item's behavior on the vehicle.
>
> c) the functionality of the item under consideration required by other items and elements.
>
> d) the functionality of other items and elements required by the item under consideration.
>
> e) the allocation and distribution of functions among the involved systems and elements; and
>
> f) the operational scenarios which impact the functionality of the item.

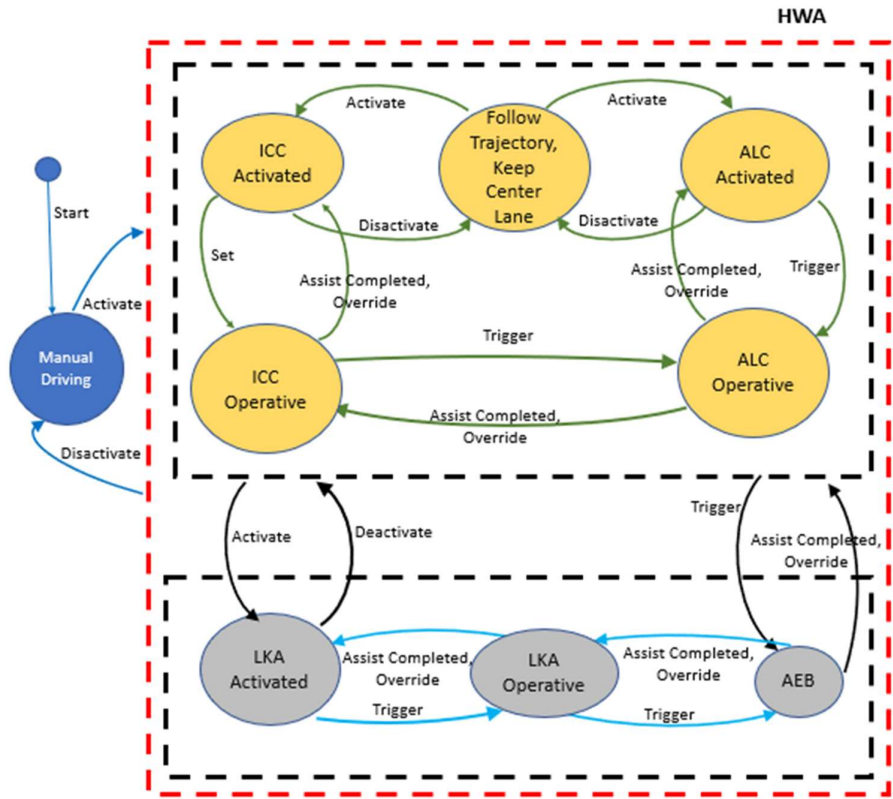

Fig 1.  HWA state diagram including other advanced assisted driving features.

One crucial piece of information that will add rigor to the above requirements of the item's boundary is the specification and details of all parameters, variables, signals, messages, value ranges, and units of signals or messages flowing at the various interfaces among the various elements internal and external to the item. Nevertheless, the standard does not explicitly specify any rigor in the above characterization of the Item, and this is left to the criteria of the developers. Being a non-prescriptive standard and not addressing rigor explicitly, ISO 26262[19] leaves plenty of room for developing an automated vehicle with an insufficient level of safety, which must be avoided at all costs.

Nevertheless, even if one includes the above pieces, it does not mean that an Item Definition has been performed in a rigorous fashion. What is missing is a set of additional data, information, analysis, or other studies to characterize the Item in a more rigorous fashion. Examples of this additional information include state transition diagrams (also called finite state machine diagrams), state transition tables, sequence diagrams, use case diagrams, and others. The diagrams mentioned above are also recommended by the International Council on Systems Engineering (INCOSE)[20] and are part of standards such as SysML and UML.

### B. Example 2: Rigor in the Performance of Functional Safety Analysis

The ISO 26262 standard requires performing a hazard analysis and risk assessment (HARA) at the concept phase in order to determine the safety risk associated with the Item under analysis. The outcomes of the HARA process are the identification of as set of safety hazards characterize by their risk levels called ASILs which could be ASIL A, through ASIL D, the latter being the most stringent. The ISO 26262 standard requires that the analysis and design of Items with the most stringent risk characterization be analyzed and design with corresponding level of rigor. In terms of hardware safety metrics, the standard requires quantitative targets listed in Table 2.

---

[19] Although we are addressing ISO 26262 specifically, the same observation is true for ISO 21448 (SOTIF) which addresses automated vehicles more explicitly.
[20] https://www.incose.org/

Table 2. Quantitative metrics of ISO 26262

| Random Hardware Metric | ASIL B | ASIL C | ASIL D |
|---|---|---|---|
| Single-point fault metric (SPFM) | ≥ 90% | ≥ 97% | ≥ 99% |
| Latent fault metric (LFM) | ≥ 60% | ≥ 80% | ≥ 90% |
| Target values for Probabilistic Metric for random Hardware Failures (PMHF) | < $10^{-7}$ $h^{-1}$ | < $10^{-7}$ $h^{-1}$ | < $10^{-8}$ $h^{-1}$ |

To take into account failure diagnostic coverage of safety mechanisms, Table 3 lists the targets of failure rate classes (FRC) of hardware parts regarding single-point faults. The failure rate corresponding to failure rate class 1 (FRC 1) shall be less than the target for ASIL D divided by 100; the failure rate corresponding to failure rate class i, i > 3 shall be less than or equal to 10(i-1) times the failure rate corresponding to failure rate class 1.

Table 3. Targets of failure rate classes (FRC) of hardware parts regarding single-point faults

| ASIL of the safety goal | Diagnostic Coverage with respect to residual faults | | | |
|---|---|---|---|---|
| | ≥ 99.9% | ≥ 99% | ≥ 90% | ≥ 90% |
| D | FRC 4 | FRC 3 | FRC 2 | FRC 1 + dedicated measures |
| C | FRC 5 | FRC 4 | FRC 3 | FRC 2 + dedicated measures |
| B | FRC 5 | FRC 4 | FRC 3 | FRC 2 |

It is up to the implementor to determine the level of rigor in the identification of hazards, their ASIL levels, and the determination of the hardware metrics based on the proposed safety mechanisms or safety measures. Clearly, the more rigor is used, the more accurate and reliable the determination of the risk and the performance of the HW metrics. In this regard, the use of appropriate tools is important to perform the safety analysis and design in a rigorous fashion.

*C. Example 3: Rigor in the Performance of SOTIF Safety Analysis*

Performing rigorous safety analysis is perhaps most needed in the context of the SOTIF standard (ISO 21448). Fig. 2 illustrates a harm model of ADS/ADS that takes into account all sources to harm that includes scenarios, misuse, performance limitations, insufficiency of specifications and others taken into account by SOTIF [19]. Our definition of safety, *absence of unreasonable risk,* is foremost a probabilistic definition because for automated vehicles, risk is probabilistic. Indeed, the level of unreasonableness can be decided if we have a probabilistic measure of risk. For the model of Fig. 2, one such probabilistic measure is the probability of harm, $P_H$ which is,

$$P_H = (P_{FS} + P_{UB}) P_{SCS} P_{IP} + P_{ODE} \qquad (1)$$

Where,

$$P_{UB} = P_{TC} ( P_{FI} + P_{SM}) \qquad (2)$$

$$P_{FI} = P_{IS} + P_{PL} \qquad (3)$$

And, $P_{FS}$ is the probability of failure, $P_{UB}$ is the is the probability of unintended behavior, $P_{SCS}$ is the probability of safety critical situations, $P_{IP}$ is the probability of involved persons contributing to harm, $P_{ODE}$ is the probability of hazardous ODD exit, $P_{TC}$ is the probability of existence of a triggering condition, $P_{FI}$ is the probability of existence of a functional insufficiency, $P_{SM}$ is the probability of misuse, $P_{IS}$ is the probability of insufficiency of specification, and $P_{PL}$ is the probability of the existence of a performance limitation.

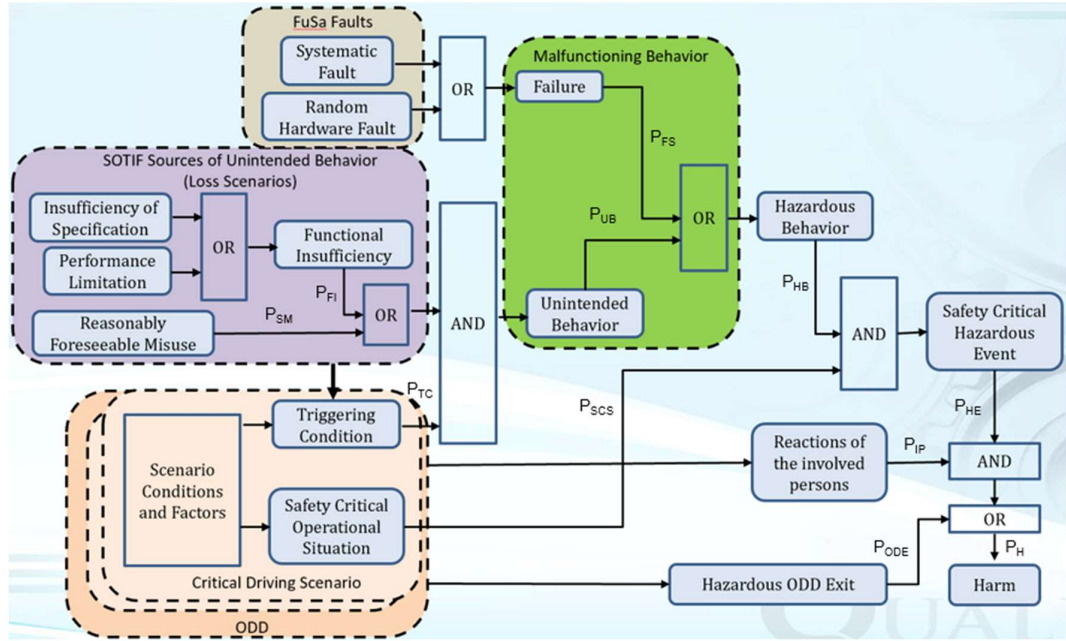

Fig. 2. Harm and accident model based on SOTIF.

Fig. 3 is a simplification of Fig. 2 where one can notice the major contributing sources to harm, namely functional safety malfunctions, SOTIF unintended behaviors, and safety critical driving scenarios including ODD. In the context of Figs. 2 and 3, rigor in SOTIF can be specified by determining quantitative measures of acceptance criteria and validation targets related to all probabilistic parameters of equations (1) through (3) such that the determination *of unreasonable risk* be established. For example, Pimentel [20] determined that, using the GAMAB principle, the probability of the existence of a performance limitation ($P_{PL}$) of the perception system to be less or equal than $10^{-8}$ failures per hour, i.e.,

$$P_{PL} \leq 10^{-8} \, h^{-1} \tag{4}$$

One can include many other examples of rigor applied to the characterization, analysis, design, and development of all aspects of an ADAS/ADS that include derivation of safety design requirements, realization of the design requirements at the system, acceptance criteria, validation targets, and requirements for hardware, software, testing, verification, validation and others. It is hoped that the above examples may be sufficient to clarify the meaning and essence of what we mean by a "rigorous analysis and design approach".

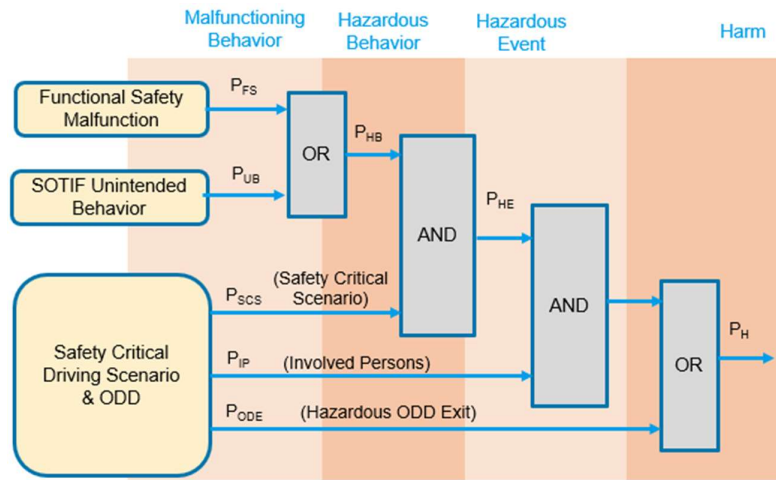

Fig. 3. Simplified harm and accident model based on SOTIF.

## 8. Question 5 - What is the role of a rigorous safety analysis and design approach? Where and when is rigor needed?

Rigor is poised to be fundamental to increase the confidence, reliance, dependability, and credibility in assuring the safety of automated vehicles. As discussed by Waymo [5], a credible and formal approach to safety is crucial for assuring safety and moving towards the acceptance of automated vehicles by all stakeholders. The role of a rigorous safety analysis and design approach is to improve the safety level of the underlying ADAS/ADS. Some rigor already exists in many works and standards, for example ISO 26262 and ISO 21448. However, we claim that it is not enough, particularly in the use of more quantitative techniques and measures. Johansson and Koopman [22] have pointed out that *"There are likely to be some aspects in which current standards are not as rigorous as they should be, resulting in higher than acceptable losses. On the other hand, there might also be a number of process measures prescribed in the standards that are too onerous for ensuring a particular level of integrity compared to alternatives. Some activities required by the standards might even have essentially no effect on safety."*

Rigor is needed in just about any activity related to the analysis, design, and deployment of automated vehicles; and this includes both, organizational processes and technical activities [23]. Rigor in technical activities has a long and rich history and is grounded in engineering and mathematical analysis and techniques where one can actually prove a certain level of safety under some assumptions [18]. Salem et al. [24] discuss a rigorous approach for an explicit representation and management of risk, e.g., behavioral specification in automated driving. In terms of rigor in organizational processes, Waymo [5] uses a definition of a safety case that leverages a number of adjectives that qualify general success criteria for a safety case: a structured argument; *a compelling, comprehensible, and valid combination of the body evidence in support of the argument*; such  qualifications are subsumed under the general notions of credibility and rigor.

There are plenty of opportunities and areas where a more rigorous approach can be used. One example is the specification of *appropriate acceptance criteria* and *validation targets* that are quantitative in nature [24]. The SOTIF standard has a few examples where a more rigorous approach can be used to evaluate the SOTIF safety risk [16]. In summary, rigor is needed in the performance of quantitative safety analysis, in the development and documentation of the safety case, and in the performance of many processes and tasks of applicable industry standards.

## 9. Question 6 - What kind of software tools support a rigorous safety analysis and design for automated vehicle development? What kind of tool categories are out there?

The sheer complexity of automated driving systems is in dire need of sophisticated and integrated software tools of all sorts to help teams to be more efficient and productive but more importantly to help ensure the safety of ADAS/ADS. The tools that we will be reviewing in this section belong to the general area of Application Lifecycle Management (ALM) tools. ALM tools enable users to work cooperatively during the product lifecycle management that includes governance, development, and maintenance. Although software ALM tools target industries such as aerospace, defense, automotive, industrial, medical devicess and semiconductors, the emphasis in this paper is automotive in general and more specifically, automated vehicles. These tools enable small or large teams, to work cooperative in company projects in an integrated and efficient manner.

One can distinguish two types of ALM tools, *stand-alone* and *cooperative*. The former are single tools that address specific tasks or problems and do not exhibit a high degree of integration with other tools while the latter are a set of comprehensive set of tools highly integrated within the set and outside of the set with tools from other companies. It is beyond the scope of this paper to discuss in a comprehensive way all ALM tools categories for the development of ADS systems and thus we will briefly discuss some tool examples.

One example of a *stand-alone tool* is Ansys Medini Analyze software[21] which is a model-based, integrated tool supporting safety analysis for safety-critical electrical and electronic (E/E) and software (SW) controlled systems. This tool implements key safety analysis methods such as hazard analysis and risk assessment (HARA), hazard and operability (HAZOP) analysis, fault tree analysis (FTA), failure modes and effects analysis (FMEA), failure modes, effects and diagnostic analysis (FMEDA) complying with standards such as ISO 26262 and SOTIF (ISO 21448). Medini integrates with other tools such as DOORS®, MATLAB®/Simulink®, Jama Software and others.

An example of a *cooperative* tool is Jama Connect® which is a Requirements Management software and Requirements Traceability solution[22]. Jama Software enables teams to manage product requirements and enable live traceability across the development process, in order to reduce cycle times and improve product quality. Unlike stand-alone tools, Jama Connect implements big portions of industry standards such as ISO 26262 dealing with requirements management. Jira Software from Attlassian is another cooperative tool that allows bug tracking, issue tracking and agile project management[23]. Jira is used by a large number of companies and users globally for project, time, requirements, task, bug, change, code, test, release, sprint management. Like Jama Connect, Jira Software implements portions of industry standards.

---

[21] https://www.ansys.com/products/safety-analysis/ansys-medini-analyze
[22] https://www.jamasoftware.com/
[23] https://www.atlassian.com/software/jira/guides/getting-started/introduction#what-is-jira-software

Another example of a cooperative tool is Simphera from dSPACE[24], a web-based tool for simulation, verification, and validation in automated vehicle development. This tool enables the user to access a myriad of realistic scenarios and run scalable simulations and it integrates with a variety of SIL and HIL platforms.

The ALM tools described so far address specific needs of organizations around specific activities such as safety analysis (Medini Analyze), requirements management and requirements traceability (Jama), issue tracking and project management (Jira), and test, verification, and validation execution (Simphera) regardless of supporting the implementation of industry standards. It so happens that some of their functionality actually help implementing some of the industry standards, but they are side benefits rather than design goals.

## 10. Tool support of a rigorous safety analysis and design framework for automated vehicle development

It is not just the tool, rather the specific quantitative metrics or other aspects of the safety analysis and design activities that would qualify a tool as rigorous. It is not entirely up to the tools to support rigor; it also depends on the specification of rigor done at the design stage which is done independently of the use of any of the tools. In terms of a pseudo mathematical expression, the above can be written as:

*Rigorous safety analysis and design = Rigorous specifications + Rigorous tools*   (5)

Thus, the degree of rigor in the specification and in the tools contribute significantly to the overall rigor in the safety analysis and design.

Not all tools support the various frameworks for automated vehicle development, much less using a rigorous safety analysis and design approach. Indeed, the tools discussed above are very general supporting several industry domains in addition to automotive and the rigorous aspects of these tools vary; however, they are not stringent.

In summary, there is a myriad of tools available that depend not only on the approach followed but also on their support of specific methodology of design such as product lifecycle, risk based, and systems engineering. In the following we describe a tool that follows a rigorous approach to safety analysis and design while implementing specific industry standards.

## 11. Introduction to the EwQIMS Functional Safety (FuSa) Module

There is a category of tools whose design goals are actually to implement processes and activities of industrial standards. One such tool package is EwQIMS from Omnex which was designed to implement many processes and activities required by automotive industry standards such as APQP[25], IATF 16949[26], ISO 26262, ISO 21448, and ISO/SAE 21434 [15,16,17]. EwQIMS[27] is a comprehensive ALM package of integrated tools developed by Omnex that supports new vehicle design activities based on the industry standards APQP, IATF, functional safety, SOTIF, and automotive security. The package enables small or large teams to work

---

[24] https://www.dspace.com/en/inc/home/learning-center/recordings/learningconnections/simphera-automated-driving-sil.cfm
[25] https://www.qad.com/blog/2023/08/the-5-phases-of-apqp
[26] https://www.iatfglobaloversight.org/iatf-169492016/about/
[27] Functional Safety Software | ISO 26262 Compliance | Automotive (omnexsystems.com)

cooperatively on organizational processes and technical design and development activities appertaining to a specific automotive product.

**A. EwQIMS Features**

As noted, simply implementing industry standards in product designs does not guarantee a safe system. We also need rigor, and this is where some tools can offer valuable essential characteristics and features. To perform its functions rigorous fashion, the most valuable essential tool characteristics are:

- Support the development of processes and activities of target standards.
- Perform quantitative hazard analysis and risk assessment to support system analysis and design activities.

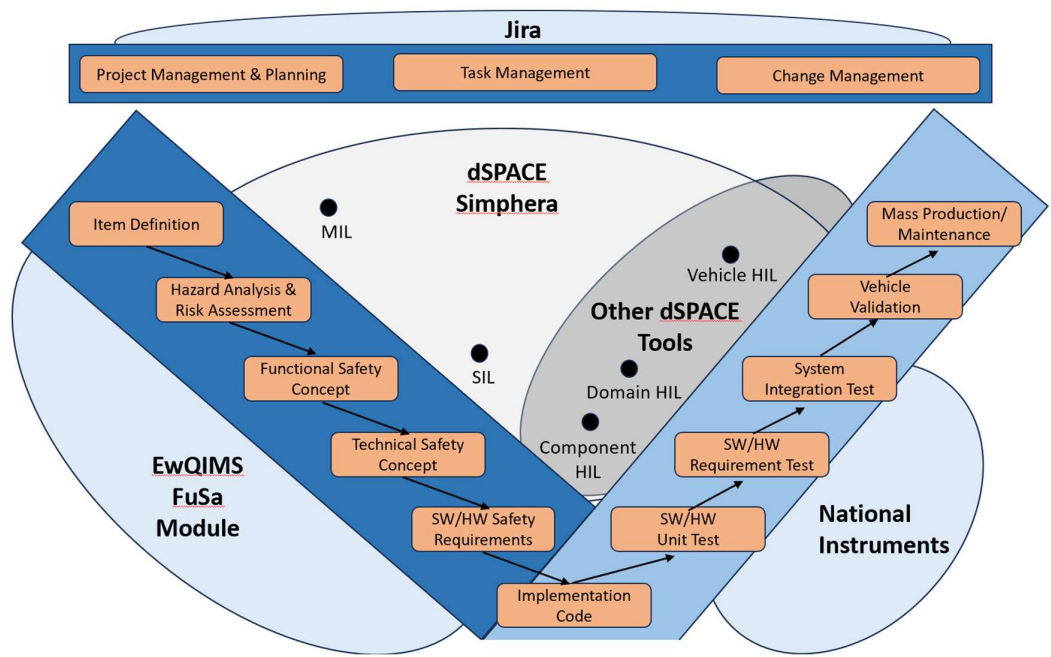

Fig. 4. Some tools that support the implementation of the ISO 26262 standard.

EwQIMS is a tool designed from the ground up to support processes and activities of industry standards. More specifically, in terms of ISO 26262, the EwQIMS FuSa module support many processes and activities of ISO 26262 including:

- Systems engineering
- Project, Safety and Risk Management
- Support of product lifecycle phases
- Concept phase including Item Definition, architectural diagrams, and HARA.
- Product design and development phase: Design tools
- Requirements modeling, elicitation, and documentation.
- Traceability management of requirements, ASIL, and tests
- Other organizational processes, such as safety management
- Tools that support testing, verification and validation activities of ISO 26262
- Functional safety work products

- Integration with other Automotive SPICE (ASPICE), Cybersecurity, SOTIF, and IATF platforms

The quantitative feature of EwQIMS consist of tool components to calculate HW random metrics SPFM, LFM, and PMHF. This complements the rigorous specification afforded by the tool to characterize the processes and activities of the target standard, i.e., ISO 26262.

As noted above, a rigorous safety analysis and design depends on rigorous specifications and rigorous tools. Some tool components of the FuSa module of EwQIMS that contribute to the overall rigorous safety analysis and design of a product include: FMEA, safety analysis, Calculation of Single-point fault metric (SPFM), Latent fault metric (LFM), Probabilistic Metric for random Hardware Failures (PMHF), and DFA. Rigor in EwQIMS is in the level of detail and comprehensiveness afforded by the various tool components and activities to be completed by its users. As depicted in Fig. 4, the FuSa module of EwQIMS fully supports the processes and activities of the left side of the V-model of the product design and development phase of ISO 26262. In addition, outputs of the EwQIMS FuSa module are compatible with other tools shown in Fig. 4 that include dSPACE (Simphera and other tools supporting MIL, SIL, and HIL), and National Instruments.

## 12. Summary and Conclusions

Although safety has been deemed paramount for the deployment of automated vehicles, the level of safety offered by commercial offerings is insufficient. Stakeholders do not agree on definitions and characterizations of what is meant by safety of automated vehicles including how to measure it and how to design for it. We use the definition of safety of UL 4600, the *absence of unreasonable safety risk.* W discussed a number of topics intrinsic to the answers to important questions raised about the safety of automated vehicles. Significant shortfalls exist in technologies for achieving the end goal of safe and secure SAE Levels 3, 4 and 5 self-driving intelligent vehicles.

Main methodologies to support the development of safe automated vehicles were summarized. In addition, we identified rigor as a significant missing requirement in the literature. Rigor is characterized as an *approach and develop systems in an extremely detailed fashion in a manner that is precise, exact, accurate, clear, compelling, comprehensive, valid, exhaustive, stringent, complete, holistic, sufficiently covered, and robust*. We also provided a discussion of rigor in the design, development, and commercialization of automated vehicles.

We discussed software tool requirements at the organizational level to support a rigorous approach for the analysis, design, and commercialization of automated vehicles. It is not enough to have rigorous tools to support a rigorous safety analysis and design process, we also need rigorous specifications. An ALM tool, EwQIMS, is introduced emphasizing the rigorous features of its functional safety module that implements much of the ISO 26262 standard. Rigorous tools such as EwQIMS help and enable the assurance of safety in the development of ADAS/ADS. Few tools that support ISO 26262 implementation are available, particularly performing a rigorous safety analysis; EwQIMS fills this void. The industry needs a more rigorous approach for safety analysis and design of automated vehicles. The offering of rigorous safety analysis in the context of industry standards as that in the FuSa module of EwQIMS is unique and significant.